\documentclass[letterpaper,fleqn,usenatbib]{mnras}

\usepackage{newtxtext,newtxmath}

\usepackage[T1]{fontenc}
\usepackage{ae,aecompl}

\usepackage{graphicx}	
\usepackage{amsmath}	
\usepackage{amssymb}	

\DeclareMathOperator{\Norm}{Normal}

\title[Automated exocomet hunting]{An automated search for transiting exocomets}

\author[GMK et al.]{
Grant M. Kennedy,$^{1,2}$\thanks{E-mail: g.kennedy@warwick.ac.uk}
Greg Hope,$^{3}$
Simon T. Hodgkin,$^{3}$
Mark C. Wyatt$^{3}$
\\
$^{1}$Department of Physics, University of Warwick, Gibbet Hill Road, Coventry, CV4 7AL, UK\\
$^{2}$Centre for Exoplanets and Habitability, University of Warwick, Gibbet Hill Road, Coventry, CV4 7AL\\
$^{3}$Institute of Astronomy, University of Cambridge, Madingley Road, Cambridge CB3 0HA, UK
}

\date{Accepted XXX. Received YYY; in original form ZZZ}

\pubyear{2015}

\begin{document}
\label{firstpage}
\pagerange{\pageref{firstpage}--\pageref{lastpage}}
\maketitle

\begin{abstract}
  This paper discusses an algorithm for detecting single transits in
  photometric time-series data. Specifically, we aim to identify
  asymmetric transits with ingress that is more rapid than egress, as
  expected for cometary bodies with a significant tail. The algorithm is
  automated, so can be applied to large samples and only a relatively
  small number of events need to be manually vetted. We applied this
  algorithm to all long cadence light curves from the \emph{Kepler}
  mission, finding 16 candidate transits with significant asymmetry, 11
  of which were found to be artefacts or symmetric transits after manual
  inspection.  Of the 5 remaining events, four are the 0.1\% depth
  events previously identified for KIC~3542116 and 11084727. We identify
  HD~182952 (KIC~8027456) as a third system showing a potential comet
  transit.  All three stars showing these events have H-R diagram
  locations consistent with $\sim$100Myr-old open cluster stars, as
  might be expected given that cometary source regions deplete with age,
  and giving credence to the comet hypothesis. If these events are part
  of the same population of events as seen for KIC~8462852, the small
  increase in detections at 0.1\% depth compared to 10\% depth suggests
  that future work should consider whether the distribution is naturally
  flat, or if comets with symmetric transits in this depth range remain
  undiscovered. Future searches relying on asymmetry should be more
  successful if they focus on larger samples and young stars, rather
  than digging further into the noise.
\end{abstract}

\begin{keywords}
  comets:general -- circumstellar matter -- planetary systems --
  stars:variables:general -- infrared: planetary systems --
  stars:individual:HD~182952.
\end{keywords}



\section{Introduction}

Comets are a well known and common component of our Solar system.  These
bodies, like the Asteroids, are planetary building blocks and are a
remnant of the processes that made the giant planets.  Comets sometimes
appear as naked-eye objects in the night sky when they pass through the
inner Solar system; while comet nuclei are relatively small and hard to
detect, their comae can be much larger and more visible, in some cases
as large as the Sun. These comae tend to have very low optical depths,
and are generally observed in scattered light.

The first evidence that comets may exist in close proximity to other
stars came from spectral observations, which found transient absorption
features towards the young star $\beta$~Pictoris
\citep{1987A&A...185..267F,2014Natur.514..462K}. Further detections have
been made towards other stars, such as HD~172555
\citep{2014A&A...561L..10K} and $\phi$~Leo \citep{2016A&A...594L...1E}.
These features change from night-to-night, have radial velocities,
accelerations, and absorption depths consistent with apparition at a few
to a few tens of stellar radii, and with models of cometary comae
\citep{1990A&A...236..202B,2018MNRAS.tmp.1412K}.  This discovery
prompted theoretical work that explored the possibility of detecting
transiting comets in broadband photometry, predicting the depths to be
of order tenths of a percent \citep{1999A&A...343..916L}. To date, none
of the stars showing variable spectral absorption have been seen to show
broadband photometric variability that might be attributed to the same
or similar events.

The detection of photometric variation has instead relied on wide-field
transit surveys, where hundreds of thousands or millions of stars are
monitored for month to decade-long periods. In particular, the detection
of deep and irregular dimming events in \emph{Kepler} data for
KIC~8462852 renewed interest in the possibility of transiting comets
\citep{2016MNRAS.457.3988B,2018MNRAS.473.5286W}. Of the many proposed
scenarios --- including Solar System-related clouds, intervening compact
objects and interstellar material, and planetary engulfment
\citep[e.g.][]{2016ApJ...829L...3W,2016ApJ...833...78M,2017A&A...600A..86N}
--- the comet family model has seen the most development, and been shown
to explain both the short and long term variation for KIC~8462852
\citep{2018MNRAS.473.5286W}.

While this star is unique among those observed by \emph{Kepler} in terms
of the depth\textbf{, shape,} and duration of the dimming events, there
is now evidence of shallower events that appear consistent with the
predictions of \citeauthor{1999A&A...343..916L}, and that have therefore
been interpreted as exocometary transits \citep{2018MNRAS.474.1453R}.

The study by \citeauthor{2018MNRAS.474.1453R} involved examining all
photometric data from the \emph{Kepler} mission by eye, looking for new
dimming events which had not been detected by the Planet Hunters citizen
scientist project \citep[e.g.][]{2012MNRAS.419.2900F} or other
searches. They found three transit events around the star KIC~3542116,
and a similar event around KIC~11084727. These events all have a similar
shape; distinctly asymmetric with a steep initial drop in flux and a
longer tail. The events for KIC~3542116 could have a period of about 92
days, but would then also require the depth of the events to be highly
variable, as more 0.1\% events should have been detected during the
\emph{Kepler} mission if the events were periodic. Further analysis of
the light curve of KIC~3542116 revealed three much shallower events of
similar shape. Based on a remarkable similarity to the predictions of
\citet{1999A&A...343..916L}, \citeauthor{2018MNRAS.474.1453R} argue that
the most likely causes of these events are exocomets.

Of the seven events discussed in \citeauthor{2018MNRAS.474.1453R}, only
four were detected in the initial visual survey. The other three were
smaller events, only visible after more detailed examination of the
KIC~3542116 light curve. In particular, determining the shapes of these
shallow events (and thus determining their cometary nature) required
target-specific analysis for subsections of the light curve near the
event of interest using Gaussian processes. Whether this type of
detrending could be applied to the entire \emph{Kepler} dataset in an
automated way is unclear.

Here, we describe the development of an automated algorithm to perform a
similar search. The primary motivation is that automated methods are
less prone to human error, are repeatable, and allow specific hypotheses
to be tested in ways that would be difficult (or unreasonable) compared
to by-eye methods (i.e. ``can you please look through those 200,000
light curves again, but this time look for feature X that we forgot to
mention last time'').  Specifically, our basic assumption here is that
the defining feature of a photometric cometary transit is an asymmetry
that is caused by the coma and tail, and a metric that quantifies this
asymmetry is central to our search.  Section \ref{sec:data} describes
the \emph{Kepler} data used, sections \ref{sec:search} and
\ref{sec:results} describe our search algorithm, its application, and
briefly discusses the results. We conclude in section \ref{sec:concl}.

\section{Obtaining Data}\label{sec:data}

The entire \emph{Kepler} dataset was downloaded from the Mikulski
Archive for Space Telescopes (MAST). In total the data comprise four
years of photometry for roughly 200,000 distinct stars. Not all stars
were observed for the full length of the mission, so for statistical
purposes we also use 150,000 as the approximate number of equivalent
stars that were observed for the full mission duration.

The flux values for the light curves come in two products, SAP\_FLUX and
PDCSAP\_FLUX \citep[see the \emph{Kepler} data processing handbook,][for
more details]{2017ksci.rept....8S}. The SAP\_FLUX (Simple Aperture
Photometry) is the detector flux within a fixed aperture.  The data are
then cleaned using the \emph{Kepler} PDC (Pre-search Data Conditioning)
module to produce the PDCSAP\_FLUX. Here we start with the conditioned
light curves, as many of the steps in the cleaning process dramatically
improve the quality of the data. Most importantly, the light curves are
de-trended with long period variations (of order several months)
removed.  The light curves are reasonably flat, making detecting
significant deviations from the mean simpler. Additionally, the cleaning
removes or corrects many artefacts, which may otherwise be detected as
transit events. While beneficial, this cleaning comes at the cost of a
few additional artefacts that are discussed below.

The light curve files contain a SAP\_QUALITY value for each data point.
These are used to indicate events, such as cosmic rays, which may affect
the reliability of the data. To detect potential transits, all points
with non-zero SAP\_QUALITY values were discarded to remove artefacts and
reduce the chance of obtaining false-positive transit detections. We
then relaxed this criterion, to include reaction wheel zero-crossings,
when quantifying transit asymmetries.

For each target we opted to process quarters of data separately, rather
than joining the light curves into a single 4-year curve. This option
was chosen because many stars in the dataset have at least one quarter
of data missing, which would make creating and processing a joined light
curve much more complicated, with little benefit.

The \emph{Kepler} data contain the flux values at constant ($\approx$30
minute) intervals, known as `cadences'. However, there are gaps in the
data. Smaller gaps of 1-2 data points are common, and are often caused
by removing short duration anomalies as described above. These short
gaps were linearly interpolated. Larger gaps are hard to accurately fill
without potentially introducing artefacts, so were treated differently
as described below. The light curves were normalised by dividing by the
mean.

\section{The search algorithm}\label{sec:search}

\subsection{Removing Periodic Noise}

Many light curves contain periodic or quasi-periodic variation. This may
be caused by intrinsic stellar variability or surface features such as
starspots.  Since the events we are searching for are not expected to
appear periodic (i.e. comet periods are typically longer than the 4-year
duration of the \emph{Kepler} mission), we attempted to detect all
significant periodic signals and subtract them from the flux to produce
a cleaned light curve.

The fastest method to detect periodic signals is to take a discrete
Fourier transform of the data. However, this suffers from two major
flaws. Firstly, Fourier methods assume that the entire light curve is a
periodic function.  This means that oscillations with non-integer
numbers of cycles per period are handled poorly, especially towards the
ends of the segments. Secondly, a Fourier transform requires that data
points be evenly spaced in time. As noted above, this is not the case
and there are some gaps in the data.  Filling these gaps using
interpolation would be required to make this method work, potentially
losing accuracy.

Instead of Fourier methods, Lomb-Scargle methods were chosen. These fix
our problems with Fourier transforms; computing a Lomb-Scargle
periodogram does not require points to be evenly spaced without gaps, so
interpolating across large gaps is not needed. It can also handle
frequencies with non-integer cycles per oscillation, which means this
method may also be more accurate.  This paper uses the \texttt{python}
implementation \citep{2015ApJ...812...18V} of the fast $O(N \log{}N)$
method developed by \citet{1989ApJ...338..277P}.

For each light curve, the Lomb-Scargle periodogram is computed. The
highest peak, corresponding to the most significant frequency is then
found. A sine wave with this frequency is fitted to the data as an
approximation to the periodic noise, which is then subtracted from the
flux. This process is repeated until no peaks with a power $>$0.05
remain in the periodogram, and the light curve is considered
``cleaned''. Removing signals with lower power did not yield
significantly cleaner light curves.

\begin{figure*}
	\centering
	\includegraphics[width=1.\linewidth]{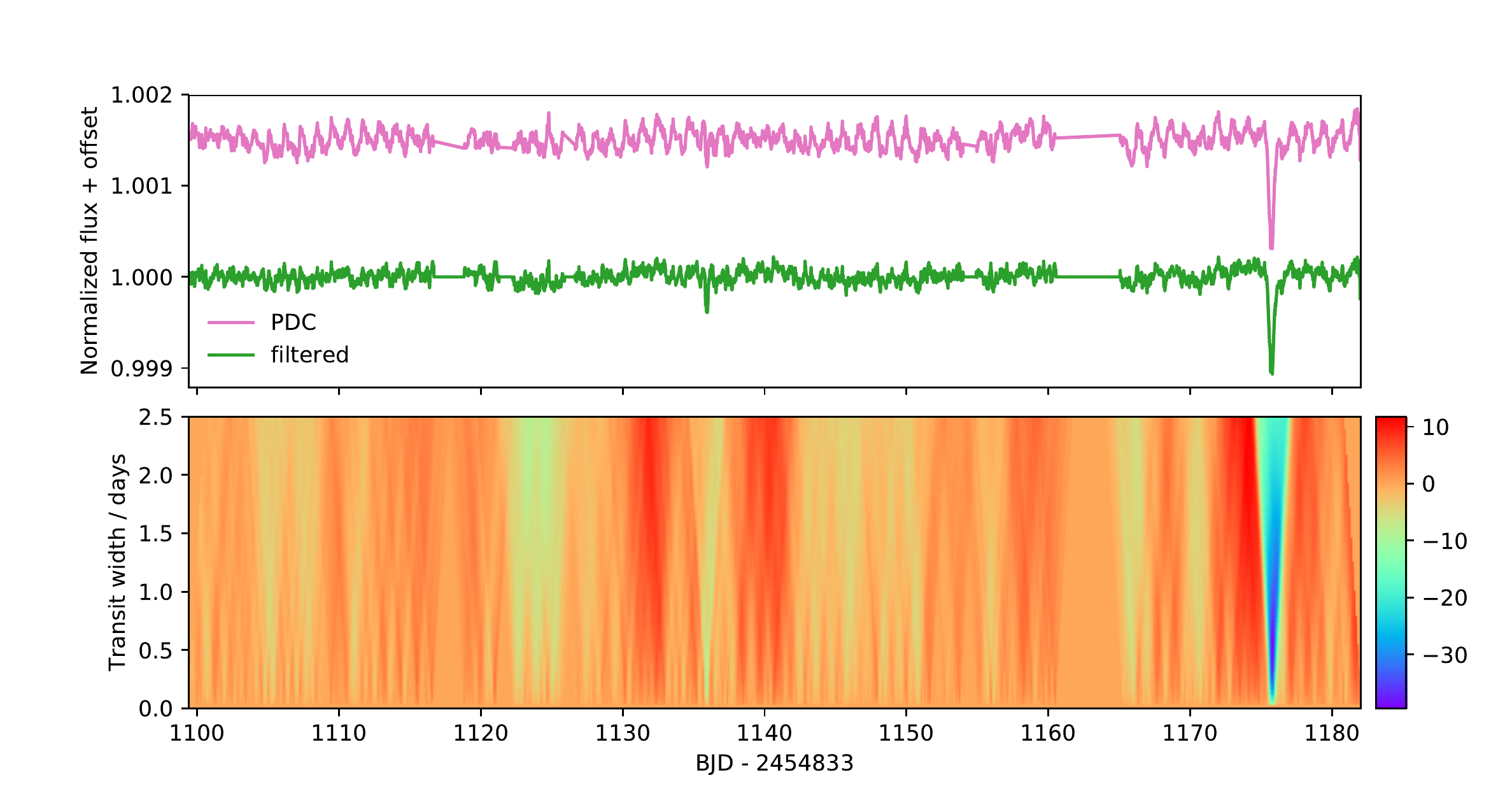}
	\caption{A segment of the raw and filtered KIC~3542116 light
          curves (upper panel), and the corresponding test statistic
          $T_{n,w}$ (lower panel).  After the periodic filtering the
          variation in the light curve has been significantly reduced,
          but the transit, which is not periodic, has been affected
          little. The minimum value is $T_{\rm tr} = -39.4$, occurring
          on day 1175, with an estimated transit width of 0.5 days.}
	\label{periodics}
\end{figure*}

Figure~\ref{periodics} shows an example of this process for KIC~3542116.
The variation has been reduced but the transit itself has not, improving
the signal to noise of the detection. Also, the cometary shape of the
transit is more visible after the processing, which will improve the
model fitting.

This process takes around 1 second per light curve, and is the slowest
part of the entire algorithm. The same algorithm using Fourier methods
would be much faster (\textless 0.1s per light curve), however it was
decided the loss of accuracy would be too large.

\subsection{Detecting Single Transits}

To detect single transits a simple box-fit approach was used. The light
curve data consists of time values $t_i$ and normalised flux values
$x_i$, $i = 1...N$. In order to keep the analysis simple, it is assumed
that the time values are evenly spaced with $t_i - t_{i-1} = \Delta t$ a
constant.  This is true for the vast majority of the data, although, as
described above, some gaps exist. These larger gaps (often of length
$\sim1$ day) had the flux set to the mean value to prevent fitting of
transits in missing data. Thus, while the width of a transit event may
be longer than found by this search if it begins or ends with a data
gap, the depth and signal to noise ratio are not affected.

In the case of Gaussian noise, the test would use a null hypothesis of
white noise with
\begin{equation}
x_i \sim \Norm(1,\sigma_x^2)
\end{equation}
where $x_i$ are all independent and $\sigma_x$ is a known value that can
be empirically measured for each light curve.  The alternative
hypothesis for a transit of depth $\mu$ centred at time $t_n$ with width
$2w\Delta t$ (i.e. $2w$ is the number of data points a transit spans) is
\begin{equation}
x_i \sim \begin{cases}
\Norm(1-\mu,\sigma_x^2) \qquad &n-w < i \leqslant n+w,\vspace{0.5em}\\
\Norm(1,\sigma_x^2) \qquad & \text{otherwise.}
\end{cases}
\end{equation}
The likelihood ratio test gives a test statistic of
\begin{equation}
T_{n,w} = \frac{(\bar{x}_{n,w}-1)\sqrt{2w}}{\sigma_x}, \qquad \bar{x}_{n,w}=\frac{1}{2w} \sum_{i=n-w+1}^{n+w} x_i
\end{equation}
where, under the null hypothesis, 
\begin{equation}
T_{n,w} \sim \Norm(0,1)
\end{equation}
for all $n, w$. A transit is indicated by a large negative value of
$T_{n,w}$ at $n_{\rm tr}$, $w_{\rm tr}$.  In practise we find that the
expectation of a normally distributed $T_{n,w}$ is not met, presumably
because residual astrophysical variations in the data violate the
assumption of Gaussian and independent $x_i$, and we therefore use the
empirical criterion described below.

This statistic can be calculated for all $n = 1 ... N$ and all $w$ up to
some maximum transit width. The maximum transit width was chosen by
looking at the durations of the events found in
\citet{2018MNRAS.474.1453R}.  The longest events found were slightly over 1
day long. We therefore use a maximum transit width of approximately
double that at 2.5 days.  With \emph{Kepler's} long cadence of
$\Delta t \approx$ 30 minutes, this duration corresponds to
$w_{max} \approx 60$. The strongest event in a given light curve occurs
at $T_{\rm tr}$, the most negative value of
$T_{n,w}$. Figure~\ref{periodics} shows $T_{n,w}$ for the filtered light
curve segment in the top panel, which shows a strong detection of the
transit occurring on day 1175.

\begin{figure}
	\centering
	\includegraphics[width=1\linewidth]{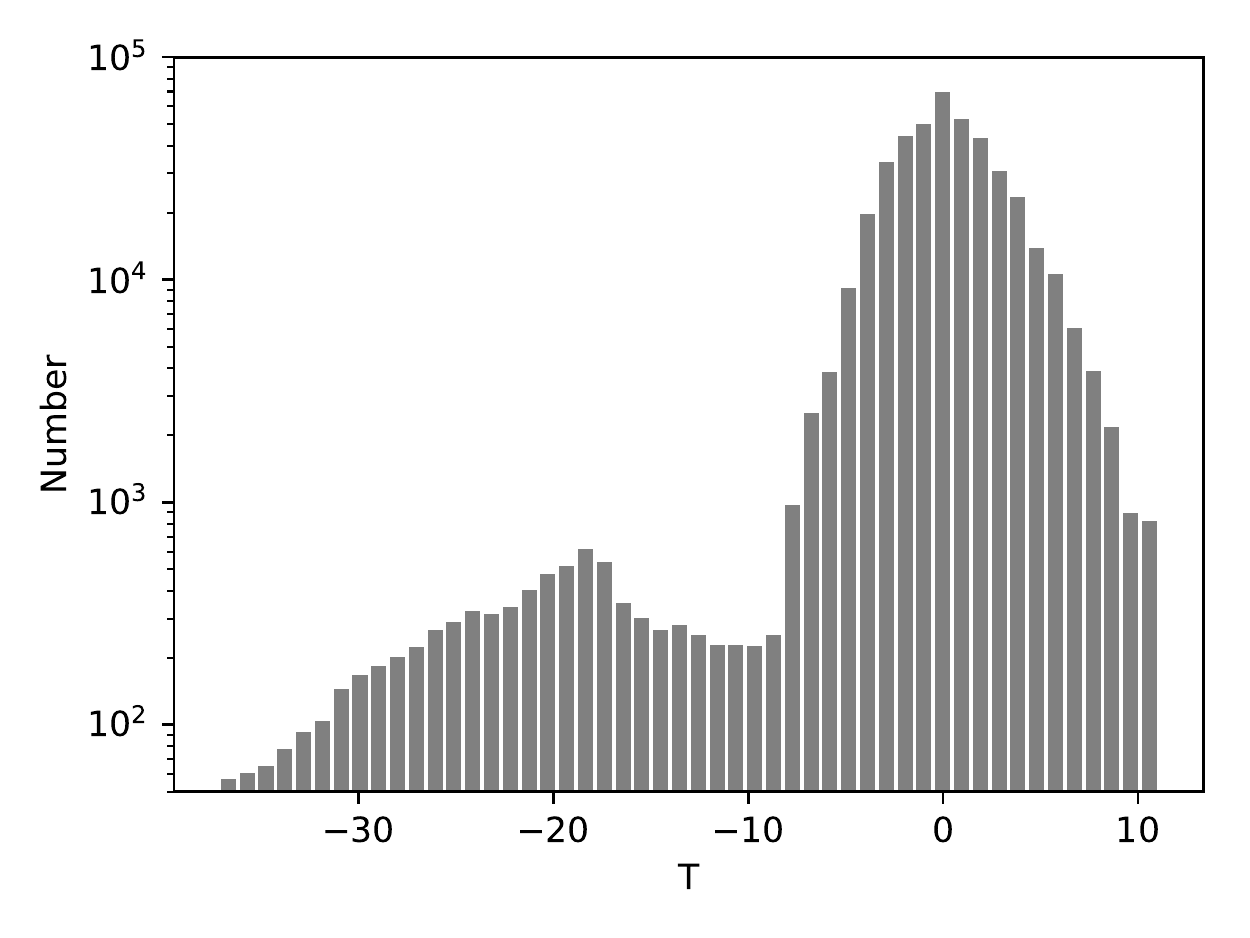}
	\caption{Distibution for transit test statistic $T_{n,w}$ from
          the light curve of KIC~3542116 shown in
          figure~\ref{periodics}. The values are centred around 0, and
          the detection at the minimum $T_{\rm tr}=-37.1$ is a strong
          outlier.}
	\label{t_dist}
\end{figure}

The next step is to calculate the signal to noise ratio (s/n) of this
transit. For this we use the distribution of $T_{n,w}$ to test how much
of an outlier the potential transit is, and compute the s/n as
$S = -min(T_{\rm tr})/\sigma(T_{n,w})$, where $\sigma(T_{n,w})$ is the
standard deviation of $T_{n,w}$ over all $n$ and $w$ considered. For the
light curve in Figure~\ref{periodics}, the distribution of $T_{n,w}$
values is shown in Figure~\ref{t_dist} (noting that this plot is shown
on a logarithmic scale, and that the tail of values below -10 is the
population of points associated with the drop in flux near day 1175).
The values of $T_{n,w}$ are clustered around 0, with a standard
deviation of $4.1$, so the signal to noise ratio for the transit is
$37.1 / 4.1 = 9$, a strong detection. While we could have used
$\sigma(T_{n,w_{\rm tr}})$ (i.e.  the standard deviation of the
$T_{n,w}$ distribution over all $n$ at $w_{\rm tr}$), we find in
practise that it makes only a small difference.

\subsection{Characterising Transit Shape}

The \emph{Kepler} data include many sources of transits that are not
comets, including exoplanets and eclipsing binary stars. In order to
distinguish these events from potential cometary transits, two models
were fitted to each transit, and the residuals compared.

Here we assume that the distinguishing feature of a comet transit
profile is its asymmetry, with a steep entry and a shallower
exit. Physically, this appearance is caused by the dust tail of the
comet, which is strongly affected by radiation forces and any stellar
wind, reducing the effective stellar mass. Particles launched from a
comet therefore find themselves on longer period orbits, which causes
them to lag behind, thus creating a tail. Whether a cometary transit
actually appears asymmetric depends on the viewing geometry, as comets
whose orbital motion near transit is largely radial have smaller tails
\citep{1999A&A...343..916L}. It is unclear whether symmetric cometary
transits should be very common; an object of fixed size is more likely
to transit near pericentre (i.e. when its motion is not largely radial),
but exocometary comae seen in calcium absorption are seen to be larger
at greater stellocentric distances \citep[where radiation pressure is
weaker, ][]{1990A&A...236..202B}, which could counteract a bias towards
detecting transits away from pericentre. In any case, as discussed by
\citet{1999A&A...343..916L} we have little means of distinguishing
symmetric cometary transits from planetary transits, so we simply
acknowledge this limitation and revisit the possibility of symmetric
cometary transits in light of our conclusions.

For a symmetrical transit, such as a planet or star transit, a simple
3-parameter Gaussian model was used.
\begin{equation}
x_{sym} = 1 - A \exp\left[- \frac{(t - t_0)^2}{2 \theta^2}\right]
\end{equation}
For an asymmetrical comet transit, a modification of this function is
used, with an exponential tail instead of a Gaussian beyond the transit
center $t_0$.
\begin{equation}
x_{comet} = \begin{cases}
1 - A  \exp\left[- \frac{(t - t_0)^2}{2 \theta^2}\right] \qquad &t \leqslant t_0,
\vspace{0.5em}\\
1 - A  \exp\left[\frac{t_0 - t}{\lambda}\right] \qquad &t > t_0.
\end{cases}
\end{equation}
The parameters for these models were optimized using the \texttt{scipy}
\texttt{curve\_fit} routine.

The fits were performed on data within a window centered on the location
of minimum $T_{n,w}$, $n_{\rm tr}$, after subtracting a linear trend on
either side of the event. A fixed number of data from the time series
array on either side of $n_{\rm tr}$ were used; for a transit
$w_{\rm tr}$ cadences (i.e. array indices) wide, data with indices in
the range $n_{\rm tr} \pm 5w_{\rm tr}$ were used. Where data gaps are
present near the transit, the time period covered by the window defined
this way is therefore greater than 10$w_{\rm tr} \Delta t$. When the
temporal window selected this way was more than 1.5 times wider than
10$w_{\rm tr} \Delta t$, the data were deemed insufficient to have
confidence in the out-of-transit baseline and/or the event itself.

The linear subtraction was necessary because a symmetric dimming event
superposed on a decreasing linear trend looks very similar to the
asymmetric events we are searching for, and local linear trends were not
always removed by the periodic noise removal.  The linear trend was
subtracted using averages of the first $3w_{\rm tr}$ points, and the
last $w_{\rm tr}$ points, under the expectation that comet-like transits
can extend farther after the transit centre than before. While this
subtraction method could in principle reduce the asymmetry of events
with particularly long tails, the shape will still appear asymmetric
(e.g. Figure~\ref{knowncomets} suggests that the asymmetry of the event
seen for KIC~11084727 could have been reduced, but this event remains
the most asymmetric among those we detected).

\begin{figure}
	\centering
    \includegraphics[width=1\linewidth]{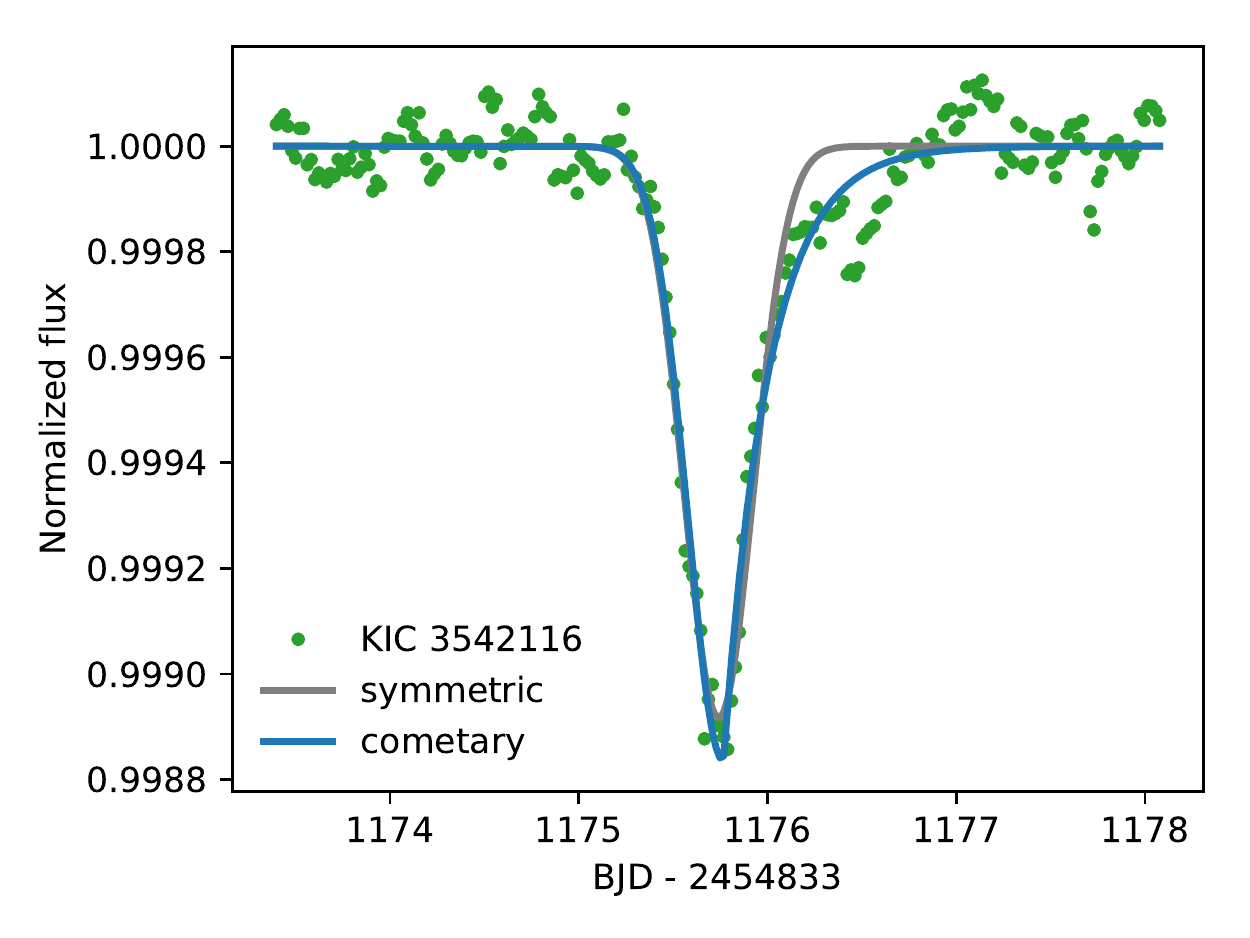}
    \caption{Symmetric and cometary models fitted to the day 1175
      transit event for KIC~3542116. The comet model provides a much
      better fit to the data. The asymmetry value for this event is
      $\alpha = 1.5$}
	\label{modelfits}
\end{figure}

Figure~\ref{modelfits} shows both models fit to a single transit of
KIC~3542116. A visual inspection shows the comet model to be a better
fit over the symmetric model in this case. This superiority can be
quantified by calculating the sum of squared residuals for each fit. We
define the asymmetry parameter $\alpha$ to be the ratio of these two
figures, that is
\begin{equation}
\alpha = \frac{\sum (x_i - x_{sym}(t_i))^2}{\sum (x_i - x_{comet}(t_i))^2}
\end{equation}
where the sum is taken over the 10$w_{\rm tr}$ window.  Values of
$\alpha>1$ denote a more asymmetric, comet-like transit.

The asymmetry parameter was calculated for a range of known objects.
Eclipsing binary stars and exoplanets typically had
$0.3 \lesssim \alpha \lesssim 1.05$. The four deeper comet events found
by \citet{2018MNRAS.474.1453R} have asymmetry values between 1.06 and 1.8.
This comparison suggests that $\alpha$ is a good parameter to determine
if a transit is likely to be a comet, with perhaps $\alpha \gtrsim 1.05$
a possible deciding characteristic.

\subsection{Filtering Artefacts}

The conditioned \emph{Kepler} data are a significant improvement over
the raw time series data, but there remain artefacts that affect our
ability to find asymmetric transits. Two main classes of artefacts were
found which produced events with profiles looking similar to a comet
transit. These would create many false positive candidates, making
finding real events difficult.  This subsection discusses these classes
and how they were detected and removed from the results.

\begin{figure}
	\centering
	\includegraphics[width=1\linewidth]{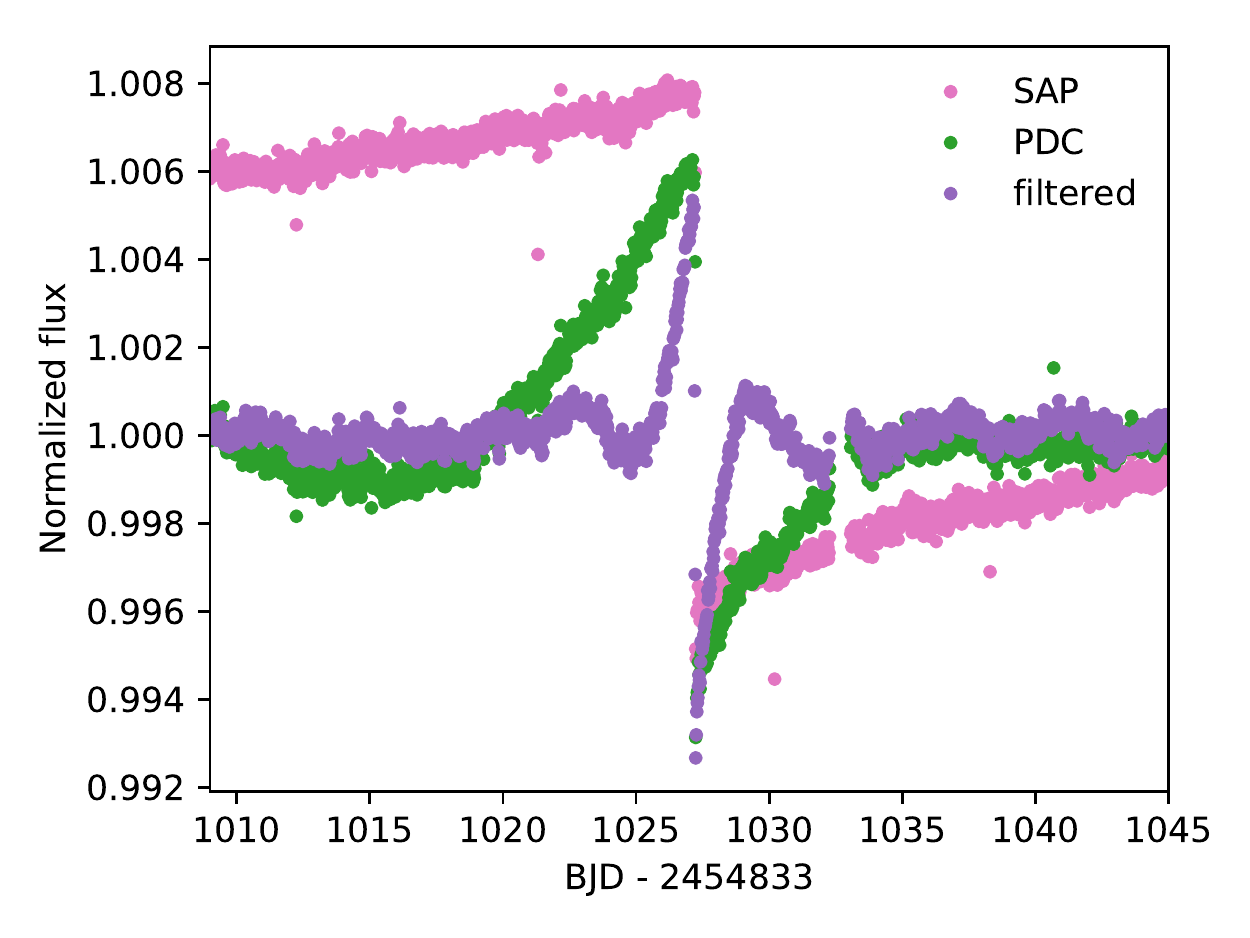}
	\caption{A light curve segment from the star KIC 5357069. The
          pink dots show the raw flux, the green dots show the processed
          flux (PDCSAP\_FLUX), and the PURPLE dots show the filtered
          flux.  There is a sharp drop in the flux occurring on day
          1027, which has not been corrected for accurately in the
          processing.}
	\label{fluxdiscont}
\end{figure}

The first class of artefacts occurs during a sudden increase or decrease
in flux level; an example is shown in Figure~\ref{fluxdiscont}. Here,
the flux drops by around 1\% in the duration of a single data point
(less than 30 minutes), and does not rise back to initial levels. The
PDC processing has attempted to fix this and flatten the curve, but has
still left a large discontinuity in the flux. In this case the
discontinuity is then exacerbated by our filtering algorithm, because it
is large.  A large number of points around day 1030 lie below the
average, triggering a transit detection by the algorithm. The algorithm
also calculates a large asymmetry for this event, with its very sharp
drop in flux and a slow rise after. Without any special handling, this
event would appear to the algorithm as a likely comet.

The solution to remove this class of events is to empirically look at
the comet entry and exit parameters, $\theta$ and $\lambda$. If one was
found to be more than three times greater than the other, or if either
is smaller than one tenth of a day, then one end of the light curve
likely has a discontinuity and the detection is considered a false
positive, so is rejected.

\begin{figure}
	\centering
	\includegraphics[width=1\linewidth]{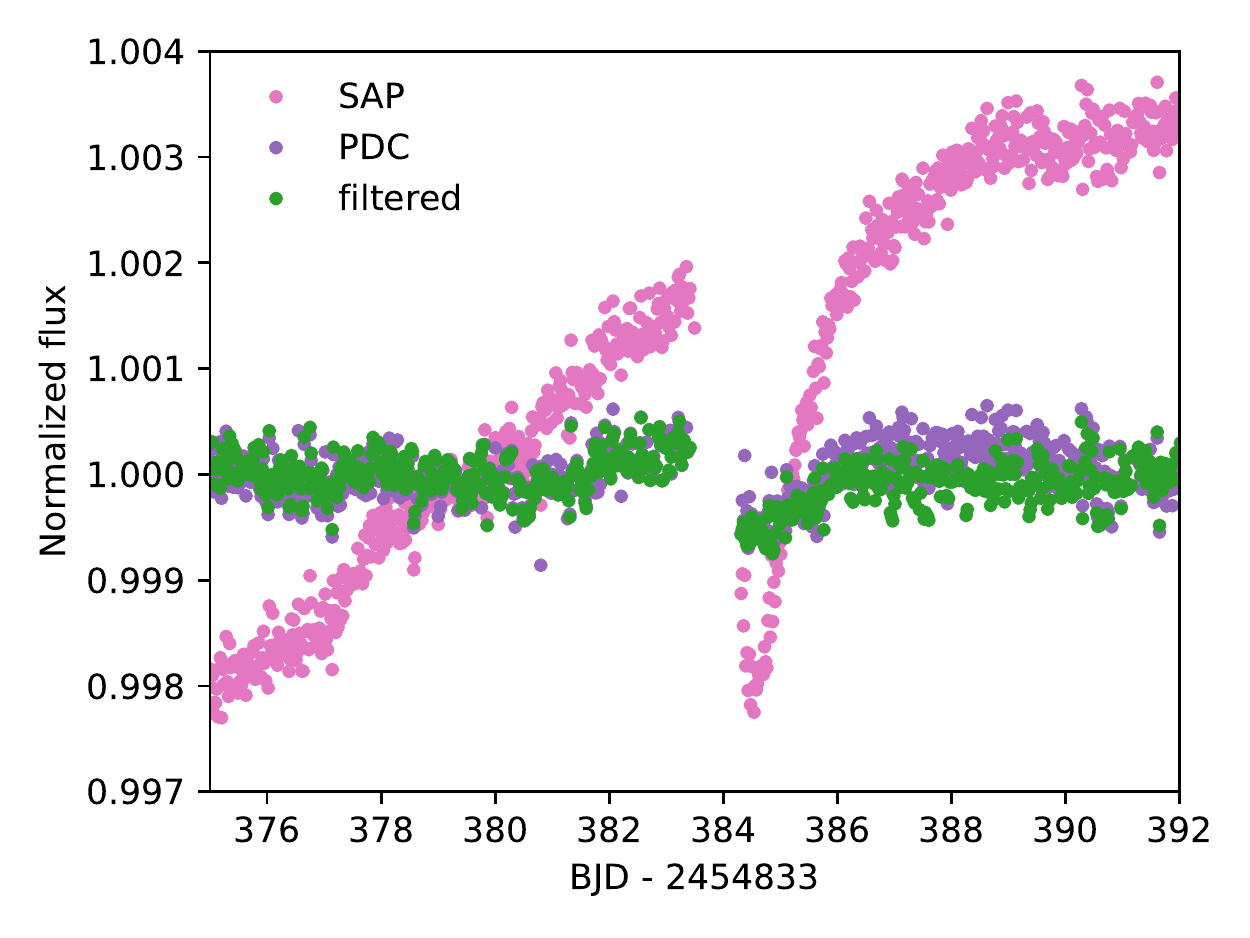}
	\caption{A comet-like artefact in the data for KIC~5985654,
          occurring directly after a short section of missing data.}
	\label{artefact}
\end{figure}

A second class of artefact common in the data is an increase or decrease
in flux directly after a segment of missing data. These occur when the
raw flux level just after and before a gap in the data do not
match. This may be related to the telescope, but could also be
astrophysical (e.g. the star is varying in brightness). These events may
not be corrected by PDC processing accurately, such as in
Figure~\ref{artefact}, but are not strongly affected by our filtering
method as the variation is typically small (i.e. the filtered light
curve looks very similar to the PDC light curve in Figure
\ref{artefact}. Mostly these are marked in the data as anomalies via the
flags, but sometimes, particularly for the very shallow dips we are
looking for, they are not. Removing these events was performed by
discarding any events where gaps longer than half a day were detected in
the two days prior to the event centre. While this criterion may discard
real transits, nearly all 5,500 putative transits removed for this
reason lie at 28 specific times spaced throughout the \emph{Kepler}
mission, and therefore any real transit that happened to occur at one of
these times would be very hard to verify.

A similar artefact to both of the above is the so-called Sudden Pixel
Sensitivity Dropout \citep[SPSD,][]{2017ksci.rept....8S}, where a light curve
can drop by $\sim$0.5\% within a single cadence. These are generally
identified and corrected by the PDC pipeline (i.e. the discontinuity is
removed), but can leave artefacts that look very similar to comet
transits. We did not attempt to filter these events automatically, and
manually remove them from the list of candidate transiting comet systems
below, either by their identification via the quality flags, or by the
presence of an obvious step in the raw light curve.

\section{Results}\label{sec:results}

The algorithm was run against the entire \emph{Kepler} long-cadence
dataset. Because the transit detection is independent of the shape
characterization, we ran the detection algorithm once, and then
experimented with the shape characterization on a subset of 67,532
transits with signal to noise ratios $S$ greater than five. While we
excluded all data with non-zero quality flags for transit detection, we
relaxed this criterion for the shape characterization; the comet transit
for KIC~3542116 at day 1176 is followed by a series of reaction wheel
zero-crossings, which can increase the noise level, and about 0.5 days
of the egress of the day 1268 event was taken while the spacecraft was
in coarse pointing. Neither of these flagged periods appear to be
accompanied by increased noise or systematic changes in the data. Coarse
pointing data are excluded by the PDC pipeline, so we do not
re-introduce those data, but we include data flagged as during
zero-crossings in the shape characterization so that the day 1176 event
is not excluded from our search. As seen below, this inclusion comes at
the cost of some false positive detections of potential comet transits.

Of the 67,532 $S>5$ transits found with all flagged data excluded,
36,948 potential transits remained after those considered too near to
gaps or light curve ends, or where model fitting failed, were
excluded. Further application of the absolute and relative width
criteria left 7,217 events. Of these, 1,336 are \emph{Kepler} Objects of
Interest (KOIs), though most are classed as false positives and are
eclipsing binaries\footnote{Using the KOI table from the NASA Exoplanet
  Archive, and the \emph{Kepler} eclipsing binary catalogue
  \citep{2016AJ....151...68K}.}. By inspecting a small number of
candidates, most of the events detected near $S=5$ are not obvious
artefacts. Some appear less plausibly significant when considered in the
context of the full quarter's light curve however (e.g.  in some cases
showing similarly large positive excursions). It is highly unlikely that
a significant fraction of these events are single transits of
long-period planets; the low transit probability would imply that these
belong to an implausibly numerous population of undetected planets, and
other searches for single transits have found tens, not thousands of
candidates \citep[e.g.][]{2015ApJ...815..127W}.

\begin{figure}
	\centering
	\includegraphics[width=1\linewidth]{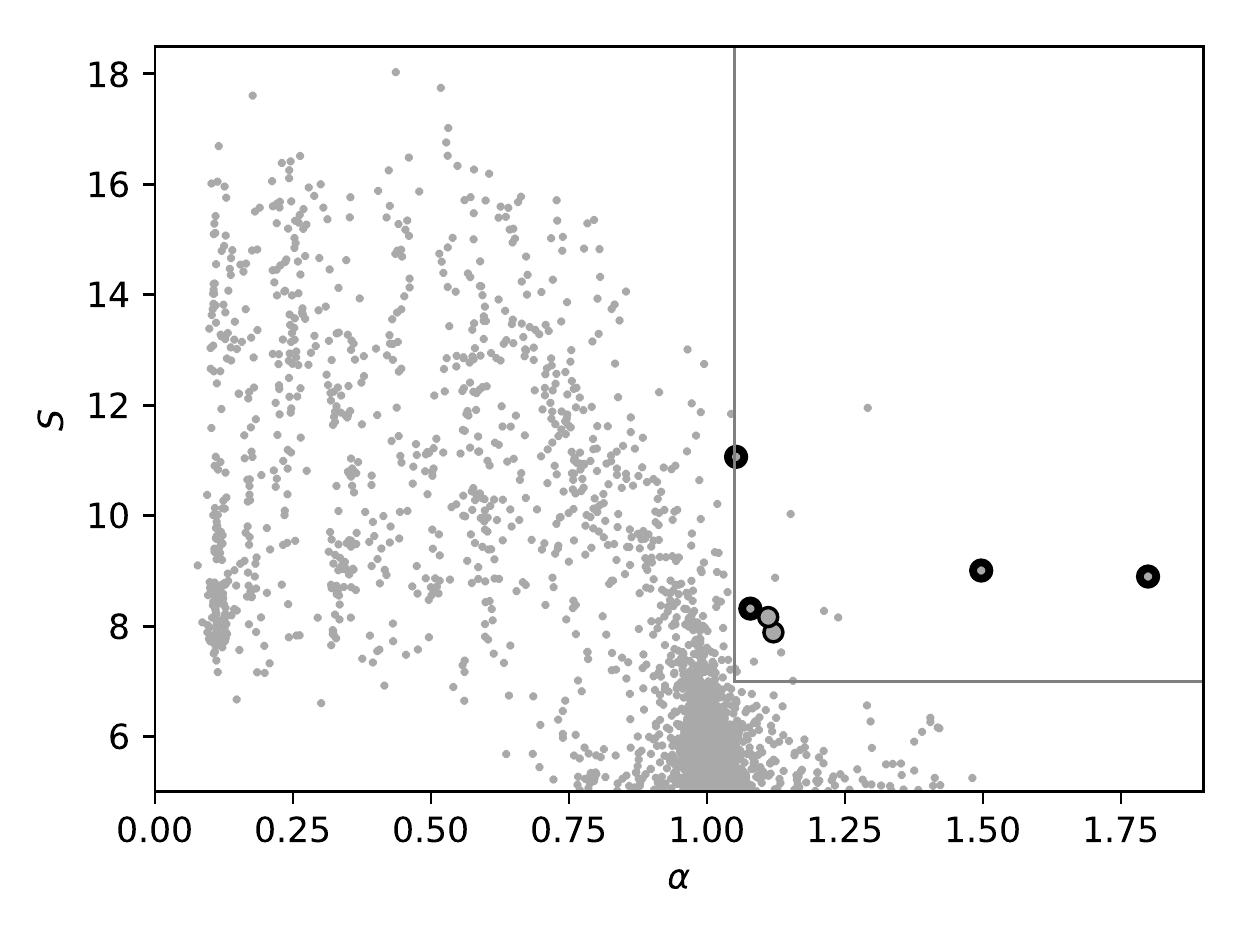}
	\caption{The signal to noise ratio $S$ against asymmetry
          $\alpha$ for detections, before manual filtering. The black
          dots correspond to the four large known events; the highest
          signal to noise ratio event to the upper left is the
          KIC~3542116 event on day 1268. The other events are, from left
          to right, KIC~3542116/d991, KIC~8027456/d1149,
          KIC~3129329/d901, KIC~3542116/d1175, and KIC~11084727/d1076.}
	\label{asym_snr}
\end{figure}

Figure~\ref{asym_snr} shows the signal to noise ratio and asymmetry for
all 7,217 detections of the algorithm with $S>5$ after automated
filtering, but before any manual filtering. The four major comet events
found by \citet{2018MNRAS.474.1453R} are marked in black on the plot, and
all lie in fairly unpopulated regions. However, these points also lie in
what may be the tail of the distribution of more symmetric transits with
$\alpha \sim 1$, or a population of candidate comet transits that are in
fact artefacts not caught by the steps described above.

To find robust transiting comet candidates, a manual examination of the
most comet-like events was therefore performed, informed by the region
in which the events reported by \citet{2018MNRAS.474.1453R} lie.  We
therefore inspected all candidates with signal to noise ratios greater
than 7 and asymmetry parameters greater than 1.05, indicated by the grey
box in figure \ref{asym_snr}. We also inspected candidates with
$\alpha>1.25$ and $5<S<7$, but found that these were either artefacts
(primarily gap-related) or cases where the fitting had failed (and that
significant asymmetry was not present).

Of the 16 events in the marked region, many (10) are false
positives. Three are SPSD events, which can be easily identified by
comparing the SAP and PDC lightcurves, and the associated quality
flags. Three are increased noise during reaction wheel zero-crossings;
these are also easily identified by inspecting the quality flags. The
behaviour in these cases appears very different to the KIC~3542116 event
on day 1176, so does not appear to call the veracity of that event into
question. One is a gap-related artefact. The remaining three are star or
planet transits, and are for stars identified as \emph{Kepler} Objects
of Interest (KOIs); one is KIC~5897826, an eclipsing hierarchical triple
system, where our search picked out two nearly overlapping consecutive
transits of different depths on day 540 (the second was shallower, thus
masquerading as a comet-like event).

\begin{figure*}
	\centering
	\includegraphics[width = 0.49\linewidth]{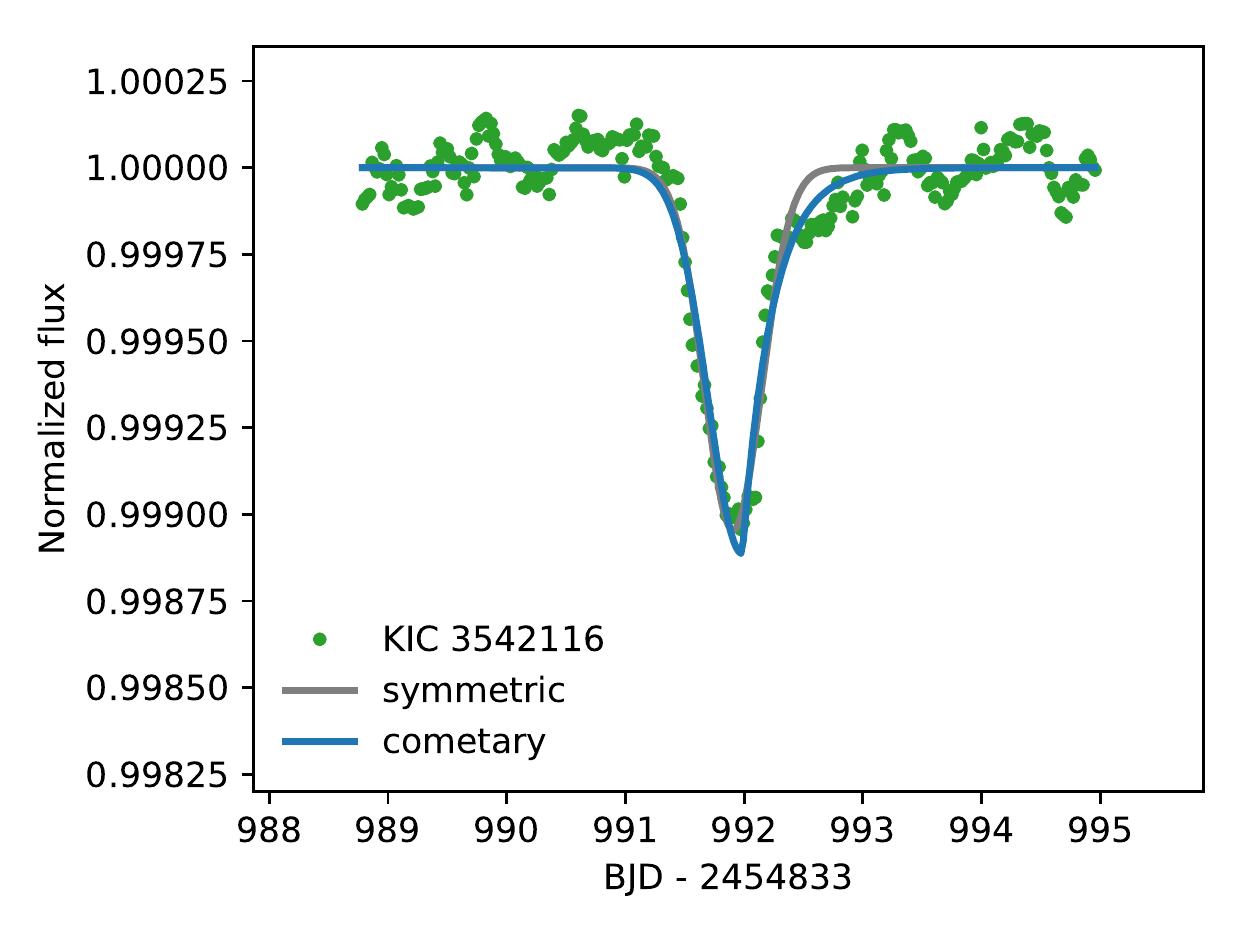}
	\includegraphics[width = 0.49\linewidth]{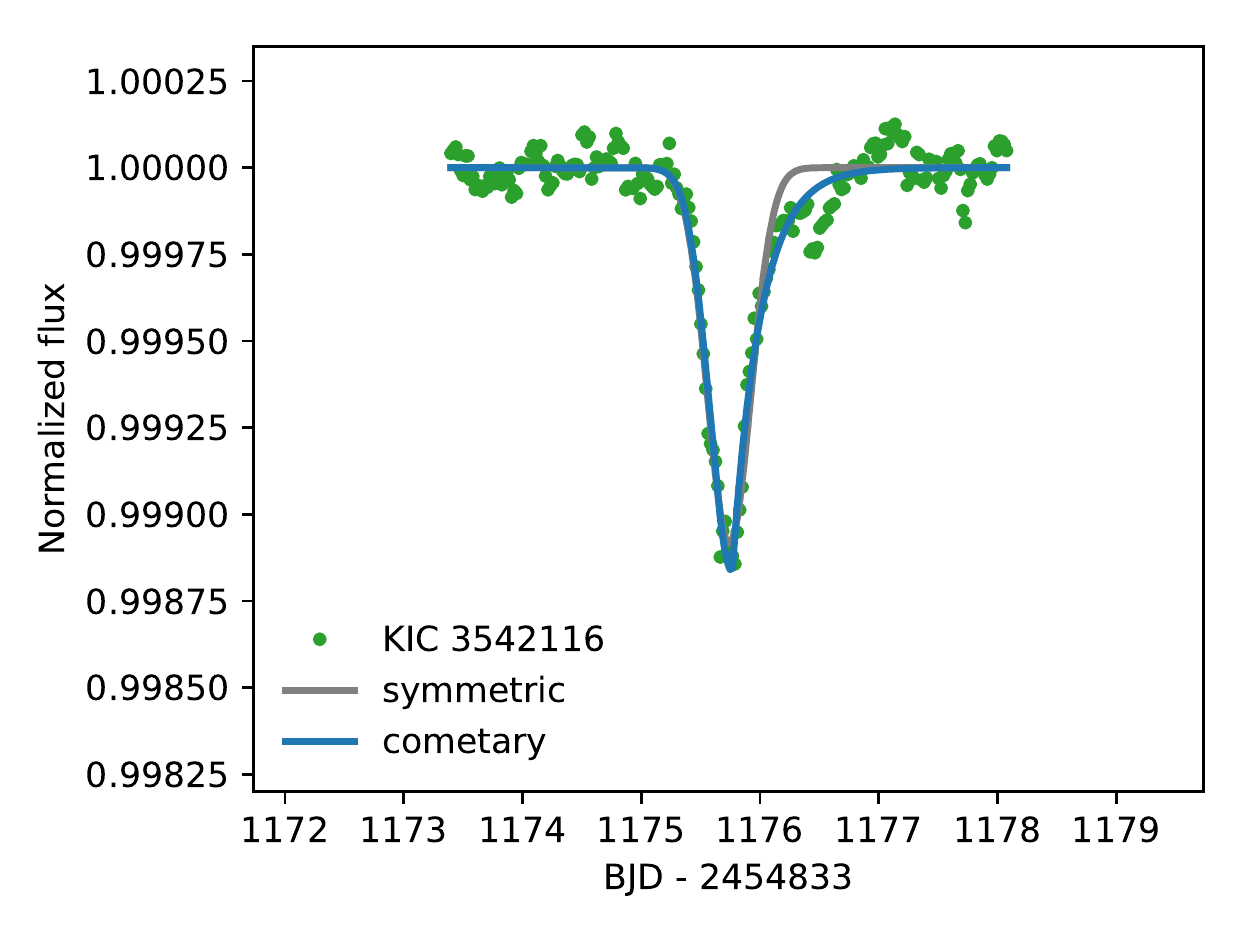}
	\hspace{0cm}
 	\includegraphics[width = 0.49\linewidth]{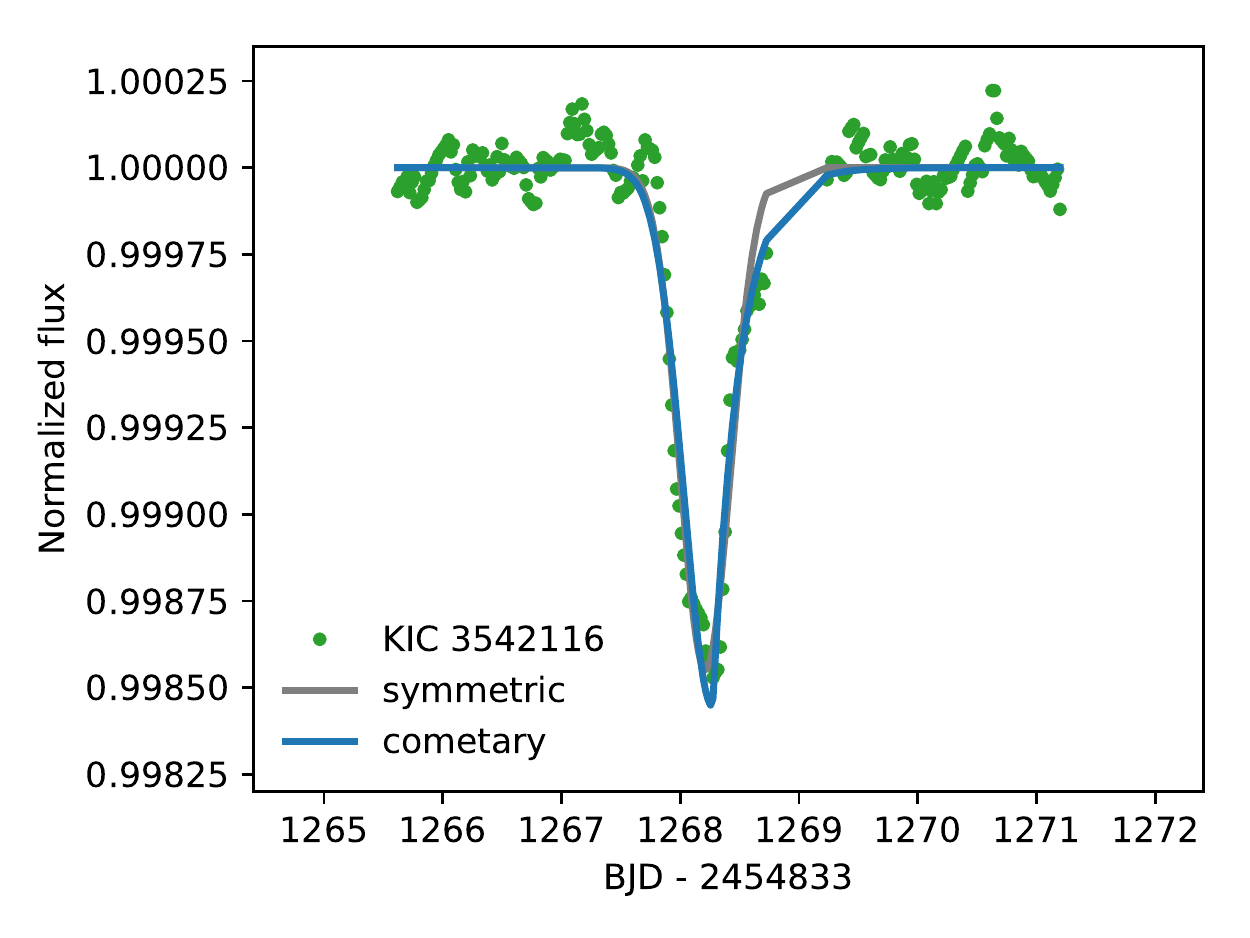}
	\includegraphics[width = 0.49\linewidth]{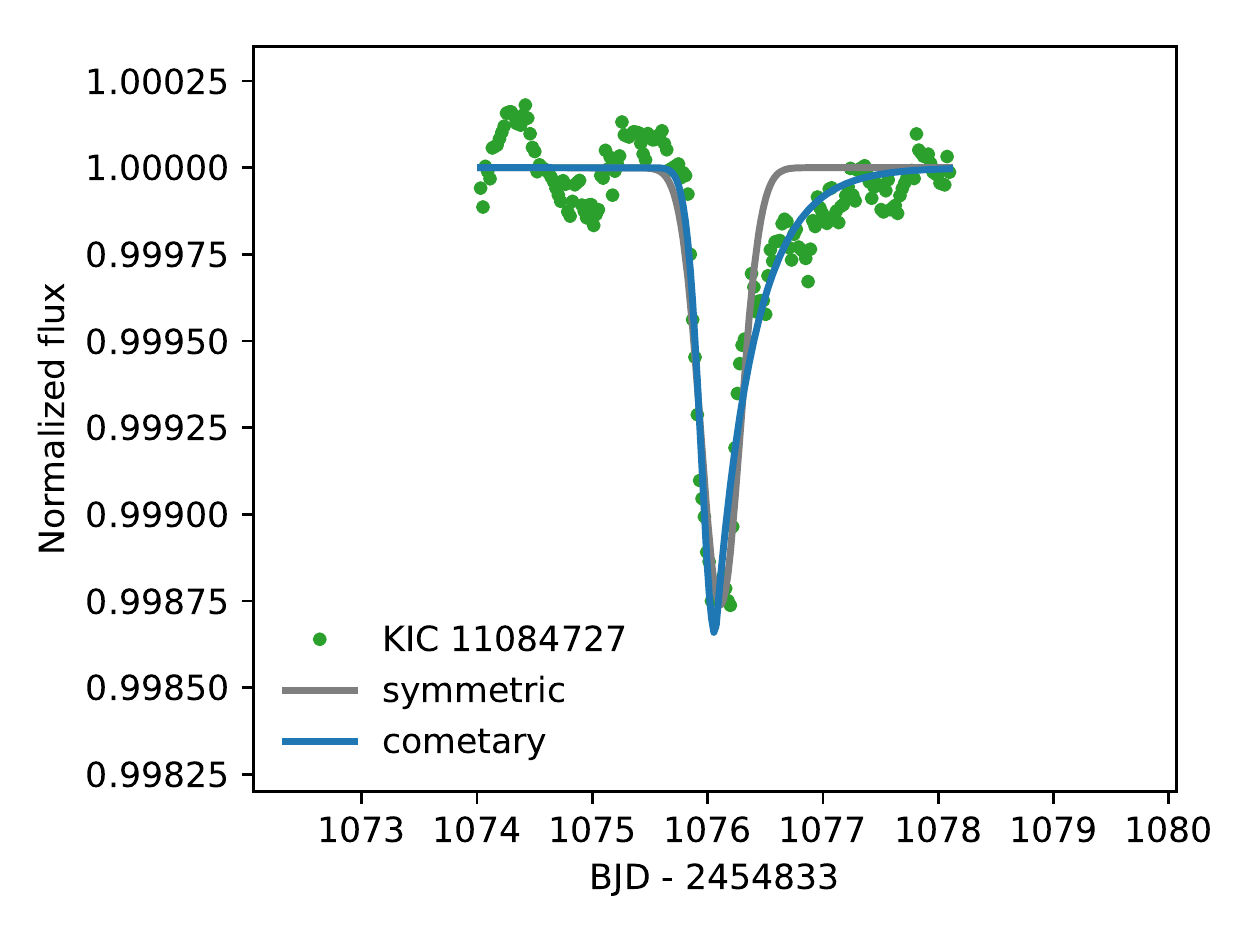}
	\caption{Redetections of the events reported by
          \citet{2018MNRAS.474.1453R}.  The data are shown within the
          $\pm$10$w$ region where the models were fit, and a linear
          trend has already been subtracted. All panels have the same
          horizontal width of 8 days, and the same vertical range.}
	\label{knowncomets}
\end{figure*}

After the manual examination, 6 events were left as non-artefact events.
These are shown in table~\ref{table}, and four are the events identified
by \citet{2018MNRAS.474.1453R}), which are shown in Figure
\ref{knowncomets}.  However, whether this recovery would have been
achieved for the day 1268 event for KIC~3542116 without prior knowledge
of its existence is debatable. The number of events with asymmetry lower
than 1.05, even at relatively high s/n, is much larger than those with
higher asymmetry. That is, the lower asymmetry bound for the search box
in Figure \ref{asym_snr} was set based on our prior knowledge of the day
1268 event, and a bound set without this knowledge might have been
slightly higher.  Figure \ref{knowncomets} shows why the asymmetry
parameter is relatively low; the spacecraft was flagged as being in
coarse point for a significant part of the egress, and is thus excluded
here. The data presented by \citet{2018MNRAS.474.1453R} suggest that the
data during this period was not affected by pointing, and therefore that
the true asymmetry for the day 1268 event is higher than 1.05.

\begin{figure*}
	\centering
	\includegraphics[width = 0.49\linewidth]{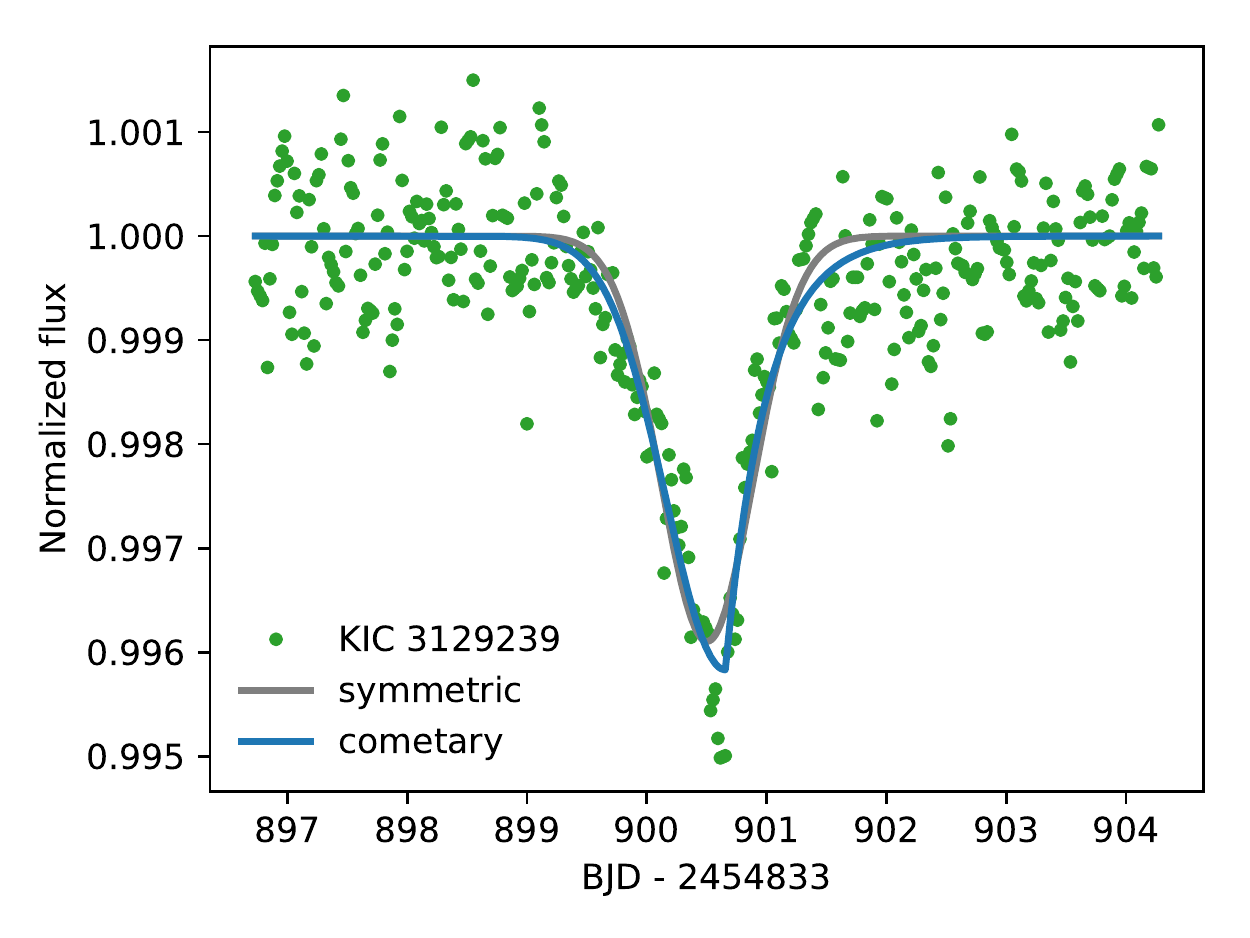}
	\includegraphics[width = 0.49\linewidth]{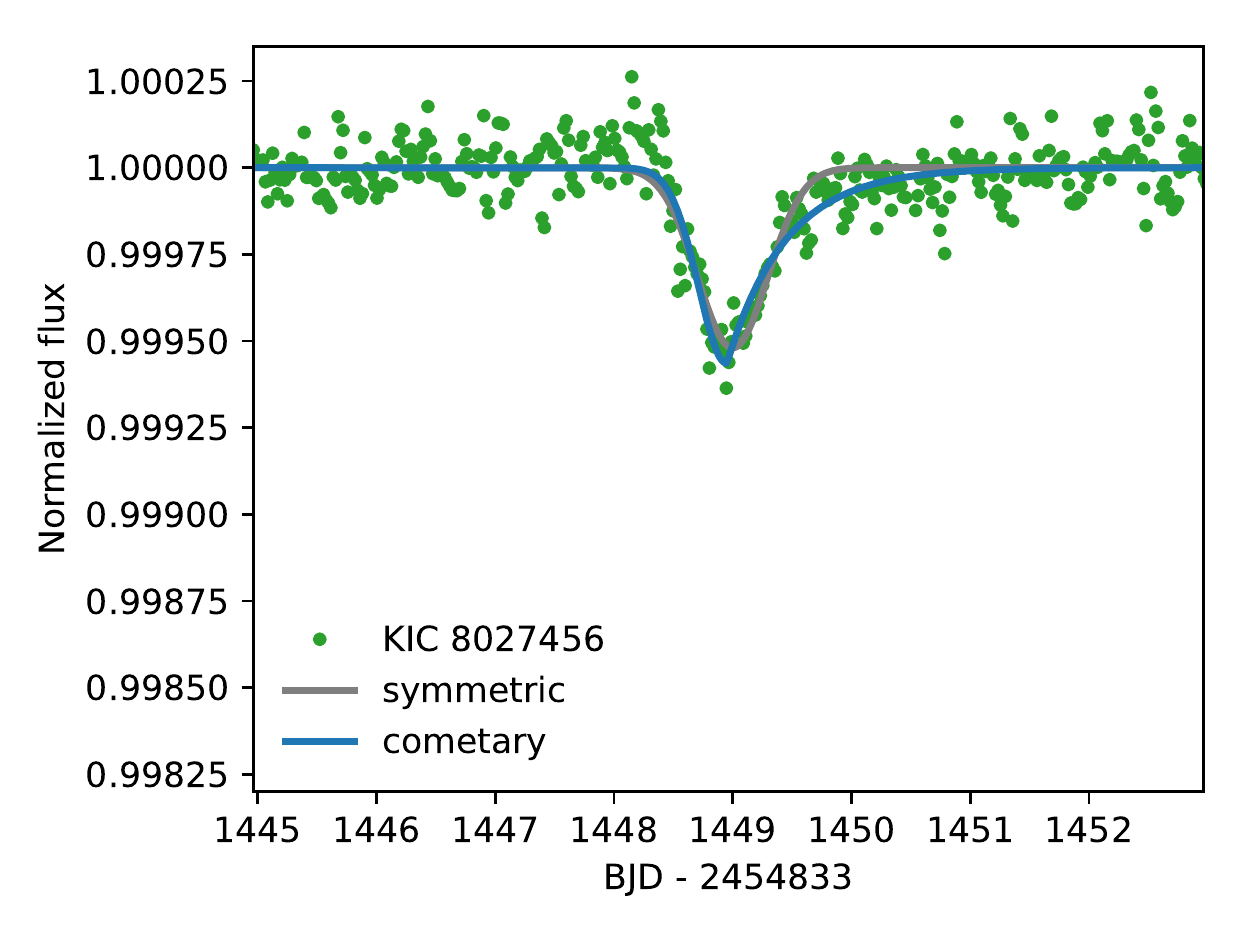}
	\caption{Two new transit detections. Each panel has the same
          width as those in Figure \ref{knowncomets}, so these events
          have longer durations. The asymmetry in the left panel is
          mostly driven by data near the transit center, so this event
          does not show the characteristic slow egress seen for the
          other events.  This event is also much deeper than all others
          identified, and the light curve has considerably higher noise.
          The asymmetry in the right panel is similar to the previously
          known events, and is shown on the same scale as the panels in
          Figure \ref{knowncomets}.}
	\label{newcomets}
\end{figure*}

The other two events are shown in Figure~\ref{newcomets}. KIC~3129239
shows an event that is several times deeper than all others that were
identified, and which visual inspection shows is less asymmetric. That
is, while the asymmetry parameter has a value of 1.12, this value arises
in part because the asymmetric model fits the data near transit center
better than the symmetric model, not because the egress appears more
gradual than ingress. Therefore, while this event appears real (i.e. is
not an artefact), it does not appear to have the characteristics of the
events being searched for, and we do not consider it further.
KIC~8027456 (HD~182952) shows a transit more consistent with the
previously identified events; the depth is about a factor of two
shallower, and the width is about 50\% greater. The shape is consistent
with our expectation of a cometary transit, so we consider this event to
be in the same class as those previously identified. The rest of the
light curve for HD~182952 is unremarkable aside from two short
($\sim$0.5d) shallow (0.25\%) dips at 277 and 282 days. Both are
associated with detector anomaly flags (the first during the dip, and
the latter a day beforehand) and this quarter's light curve has several
step discontinuities of similar magnitude. We therefore do not consider
these events further.

\begin{table}
  \caption{Comet like events in the search region which did not appear
    to be artefacts and were not in the \emph{Kepler} Objects of
    Interest catalogue. The date given is BJD - 2454833.}\label{table}
\begin{tabular}{rrrcccc}
\hline
\emph{Kepler} ID & Date & SNR & Asym. & $\theta$ & $\lambda$ & Depth \\
\hline
3542116 & 1268.2 & 11.0 & 1.05 & 0.24 & 0.23 & 0.00103 \\
3542116 & 991.9  & 8.3 & 1.08 & 0.28 & 0.25 & 0.00074 \\
8027456 & 1448.9 & 8.2 & 1.11 & 0.24 & 0.50 & 0.00035 \\
3129239 & 900.5  & 7.9 & 1.12 & 0.49 & 0.35 & 0.00288 \\
3542116 & 1175.7 & 9.0 & 1.50 & 0.19 & 0.24 & 0.00076 \\
11084727& 1076.1 & 8.9 & 1.80  & 0.12 & 0.34 & 0.00096 \\
\hline
\end{tabular}
\end{table}

Considering HD~182952 in a little more detail, the effective temperature
is reported in the range 8900 to 9800K
\citep{2015MNRAS.450.2764N,2016A&A...594A..39F}, making it hotter than both
KIC~3542116 and 11084727, which are approximately 6800K
\citep{2018MNRAS.474.1453R}. By fitting stellar photospheric models to the
available photometry, we conclude that HD~182952 shows no evidence for
an infrared excess; with only WISE observations however, the limits on
potential cometary source regions are poor. As was concluded for
KIC~8462852, the lack of IR excess is easily consistent with the levels
expected if dust is the cause of the transit event \citep[and in any
case the WISE observations occurred several years before the event,
see][for a discussion of the infra-red light curves of transiting
dust]{2018MNRAS.473.5286W}.  However, this conclusion relies on the
assumption that stars showing such events are being viewed from a
direction that places a family of comet orbits along the line of sight
to the star. If the comet orbits are in fact random and could be
detected by any observer \citep[as may be implied by the different
orbital properties of the deep and shallow events for
KIC~3542116,][]{2018MNRAS.474.1453R}, then the lack of IR excess may
provide useful constraints (but such an analysis is not the goal of this
paper).

The transit event for HD~182952 has a similar ingress parameter to the
other events, but the egress is somewhat slower. While it seems probable
that such events show a natural variation that depends on cometary
activity and orbital parameters, it may be that comets around more
luminous stars have longer dust tails, possibly caused by higher
cometary mass loss rates.

\begin{figure*}
	\centering
	\includegraphics[width = 0.49\textwidth]{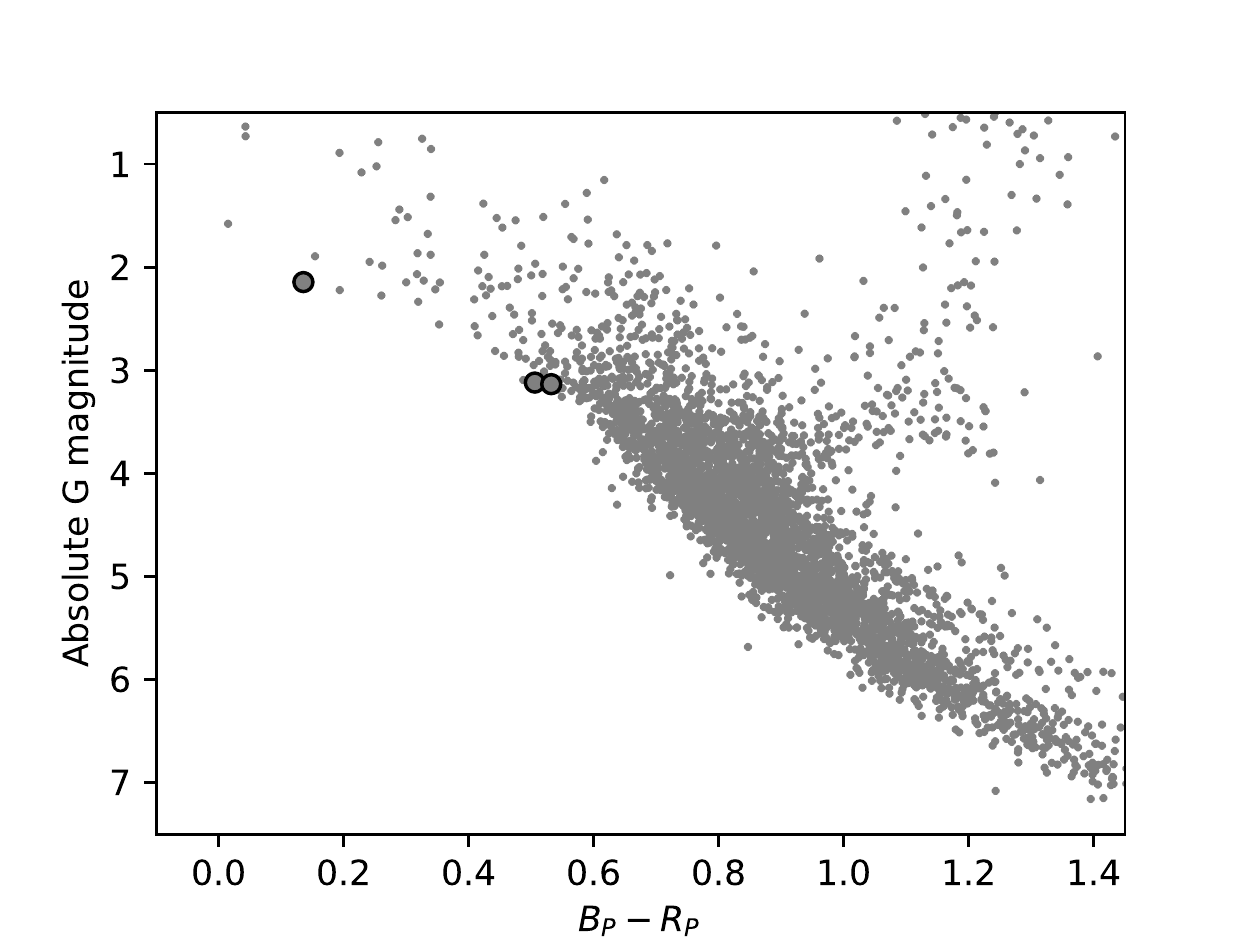}
	\includegraphics[width = 0.49\textwidth]{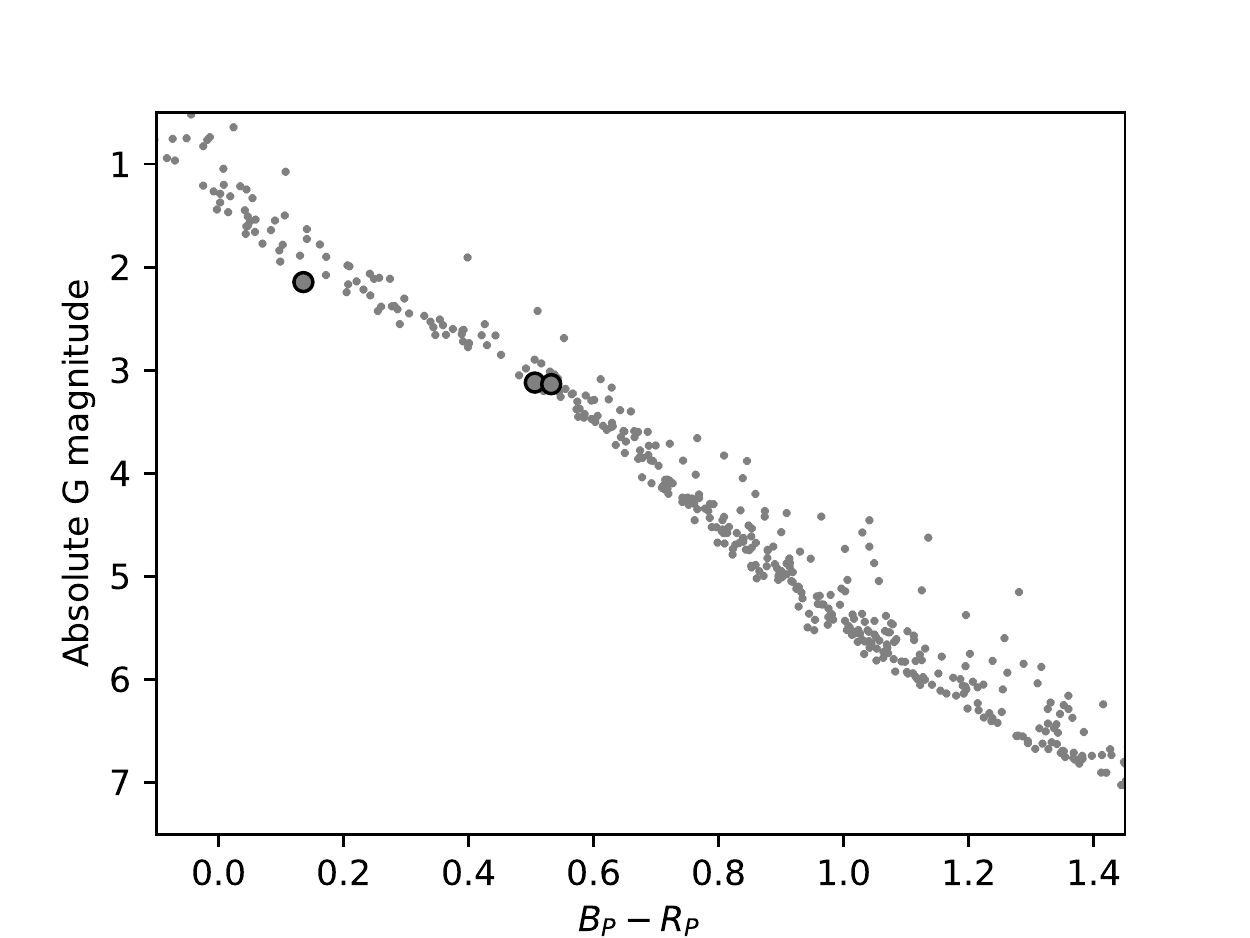}
	\caption{Gaia H-R diagrams for KIC~3542116, 11084727, and
          HD~182952 (large dots). The small dots show sets of comparison
          stars.  The left panel uses stars within 1$^\circ$ of the
          three of interest, and that have parallax$>1$, parallax
          s/n$>50$, and $G<15$ (i.e. representative stars in the
          \emph{Kepler} field). The right panel uses stars in IC~2391,
          IC~2602, NGC~2451, and the Pleiades, as a set of stars
          approximately 100Myr old. All three stars of interest lie at
          the lower edge of the distribution in the left panel, and near
          the locus of stars in the right panel, in both cases
          suggesting that they are young.}
	\label{hr}
\end{figure*}

Finally, our search does not recover the much deeper dimming events seen
towards KIC~8462852, for several reasons. Our periodic noise removal
fails badly for this star, because it assumes that any dimming events
are sufficiently shallow that they do not affect the sinusoids that are
fitted and subtracted. Periodic signals are therefore introduced, rather
than removed, and the dimming events are affected. Also, the ``D800''
dimming event is well known to show the opposite asymmetry to that
expected for a comet tail, and the family of ``D1500'' events is too
complex to be reasonably explained by single-comet models
\citep{2016ApJ...819L..34B}. Otherwise, the main implication is that
when considering how often events of a given depth occur, we must assume
that other searches would have found any asymmetric events that were
deep enough for our search method to fail. The lack of such seems very
likely to be real, as i) a search for $\ge$5\% events was performed by
\citet{2016MNRAS.457.3988B} to look for stars similar to KIC~8462852 and
none were found, and ii) any deeper comet-like events should have been
discovered in the by-eye search described by
\citet{2018MNRAS.474.1453R}.

\subsection{Evidence for stellar youth}

Figure \ref{hr} shows H-R diagrams for our three stars showing potential
comet transits, compared to stars nearby on the sky (left panel) and
young stars in nearby open clusters (right panel). The stars nearby on
the sky were selected to be within 1$^\circ$ of any of these three stars
from Gaia DR2, and restricted to have parallax $>$1~mas, parallax
s/n$>50$, and apparent $G$ magnitude $<15$ (criteria that encompass the
three of interest, and are therefore representative of the stars in
\emph{Kepler} field). The \emph{Kepler} field is near the Galactic
plane, so this comparison allows for some spread in both the colour and
magnitude of stars that is expected to arise from reddening.  In the
left panel our three stars of interest all fall at the lower envelope of
the distribution, suggesting that these stars are young. The young
clusters IC~2391, IC~2602, NGC~2451, and the Pleiades, were selected
from \citet{2018A&A...616A..10G} as they have similar ages near
100Myr.  In the left panel the stars with potential comet transits lie
very close to the locus of these young stars, suggesting that their ages
are not significantly greater than 100Myr.

It is well known that young nearby stars are more likely than old stars
to be seen to host the bright comet reservoirs known as `debris disks'
\citep[e.g.][]{2005ApJ...620.1010R}. Moreover, most stars that show
spectral signatures of transiting comets (e.g. $\beta$~Pictoris,
HD~172555) are also young. Thus, if the transiting comet hypothesis is
correct for the systems identified here and by \citet{2018MNRAS.474.1453R},
then it is not surprising that these systems should also appear to be
young. Indeed, the comparison in Figure \ref{hr} provides circumstantial
evidence that this interpretation is correct.

\subsection{Exocomet properties}

Given that the evidence for stellar youth supports the exocomet
hypothesis, we briefly consider the properties of these putative
transiting comet systems. First, we can use the transit duration to
estimate the stellocentric distance at transit $r_t$.  Rearranging
equation (21) of \citet{2018MNRAS.473.5286W} yields
\begin{equation}
	r_t < 27 M_\star R_\star^{-2} \theta^2 \, ,
\end{equation}
where $r_t$ is in units of au, $M_\star$ is stellar mass in $M_\odot$,
$R_\star$ is stellar radius in $R_\odot$ and $\theta$ is the ingress
parameter from Table \ref{table} in days. This equation uses $2\theta$
as an estimate of the ``true'' transit duration, based on the
expectation that the ingress is not affected by the comet tail. The
inequality arises because the transit duration sets the necessary
transverse velocity of the body across the face of the star, but the
velocity of an eccentric orbiting body varies around the (unknown)
orbit. The highest velocity occurs at the pericentre of a high
eccentricity orbit, which sets the maximum $r_t$, and objects with lower
eccentricity or that transit away from pericentre must transit at
smaller $r_t$ to have the same transverse velocity.

Assuming $M_\star=1.5 M_\odot$, $R_\star=1.5 R_\star$, and $\theta=0.25$
days for KIC~3542116 yields an upper limit of $r_t \lesssim 1$au,
equivalent to a period of $<$300~days for a circular orbit.  This limit
is compatible with the possible 92-day periodicity noted by
\citet{2018MNRAS.474.1453R}, but requires the comet activity to be highly
variable as only three events are seen during the \emph{Kepler}
mission. If, on the other hand, the events are unrelated, then the
orbits must be eccentric with the transits occurring near pericentre in
order for them to not repeat elsewhere during the \emph{Kepler}
mission. Assuming the same stellar properties for KIC~11084727 and
$\theta=0.12$ days yields $r_t<0.26$~au, equivalent to a period of $<$40
days for a circular orbit. This event was not seen to repeat during the
\emph{Kepler} mission, again probably requiring observation of a transit
near pericentre of a high eccentricity orbit, or variable activity.

For KIC~8027456, assuming $M_\star= 2 M_\odot$, $R_\star=1.5 R_\star$,
and $\theta=0.24$ days yields an upper limit of $r_t<1.4$au, and a
circular period of 1.1 years. The lack of repeat events points to
behaviour similar to KIC~11084727, though the possibility of fewer
events during the \emph{Kepler} mission makes this requirement less
stringent. While the orbital constraints are not strong, the duration of
the events again likely requires eccentric orbits, consistent with the
exocomet hypothesis.

\subsection{Distribution of dimming event depths}\label{ss:depth}

Finally, we consider the statistics of such events. Observed phenomena
commonly have a distribution of properties, and the most extreme are
generally detected first. Thus, it is reasonable to expect that the
dimming events seen for KIC~8462852 are the most extreme and rare of
some population \citep{2018MNRAS.473.5286W}, which also includes the
shallower events that are the focus of this paper.

In terms of the physical origin this assertion is not necessarily
secure, as it remains uncertain whether the deep events seen for
KIC~8462852 and the shallower events are the result of the same or
similar phenomena, particularly given significant differences in transit
characteristics. That is, while both have been interpreted as
exocometary transits, the evidence that led to these conclusions was
different in each case. Here, we will assume that both are drawn from
the same population in order to derive a depth distribution, but this
population might need to have a fairly broad definition, such as
``circumstellar material''.

Regardless, we may consider the depth distribution of asymmetric transit
events seen towards \emph{Kepler} stars in purely phenomenological
terms. The main caveat is that if the deep and shallow events are
physically unrelated, this distribution may not be as smooth as one
might expect from a single population. For example, the shallow events
might in fact be the most extreme exocometary events, and KIC~8462852
has a partially or entirely different origin.

Taking the total number of stars observed as 150,000\footnote{Not all
  stars were observed in all quarters; 112,046 were observed in 17
  quarters, and 35,650 were observed in 14 quarters
  \citep{2016AJ....152..158T}. Given that the transit events being
  considered here are rare (i.e. may only appear in one quarter's
  observation for a given star), we take 150,000 as an approximate
  number of equivalent stars that were observed for the entire
  \emph{Kepler} mission}, and assuming that all comet-like events deeper
than 0.1\% have been detected, the fraction of systems showing events
deeper than 10\% (i.e. KIC~8462852) is 1/150,000 ($7 \times 10^{-6}$).
The fraction showing events deeper than 0.1\% is 4/150,000
($3 \times 10^{-5}$), though this number must be considered approximate
given that not all stars have the same noise properties, and that we
have not carried out injection tests to verify our completeness.
Comparing these numbers to Figure 14 of \citet{2018MNRAS.473.5286W}, the
improved sensitivity (via by-eye and automated searches) has not yielded
as many new detections as suggested by that model (20 at 0.1\%). The
predictions for comet detections around Sun-like stars made by
\citet{1999A&A...343..916L} are yet higher; equivalent to 120 detections
at 0.1\% for the \emph{Kepler} mission.

Despite the small numbers, we might conclude that the distribution of
cometary transit depths is relatively flat, but this conclusion relies
on all comet transits being asymmetric (and that all events are
successfully identified by our algorithm). However, as Figure
\ref{asym_snr} shows, most events are symmetric, and given that most
cannot be attributed to binary and planet transits, some may be isolated
events attributable to comets, and our detection rate therefore
underestimated. Our criteria for rejecting false-positives may also be
too conservative, meaning that we have missed asymmetric transits, but
the consistency of our results with \citet{2018MNRAS.474.1453R} suggests
that we are not likely to have missed many at depths $>$0.1\%.

Thus, there are various directions for future searches for comet
transits. One would be to focus on distinguishing symmetric cometary
transits from binary, planet, and other symmetric transits.  Given the
large number of symmetric transits in Figure \ref{asym_snr}, this
approach may be fruitful, but would need to rely on properties such as
irregularity of transit timing, or transit durations that are
incompatible with non-repetition of similar events unless the orbital
eccentricity is very high. Theoretical work might also consider whether
a flat distribution of cometary transit depths is reasonable and/or
yields any insight into the comet population. The distinction between
naturally symmetric transits and symmetric comet transits may be hard to
make, so the obvious way to increase the number of detections,
particularly given the small increase found here, is to increase the
number of stars searched, and to focus on young stars where
possible. Both K2 and TESS provide observations of stars that can help
with this goal.

\section{Summary and Conclusion}\label{sec:concl}

This work presents an algorithm designed to search light curves for
asymmetric transits that could arise from transits of cometary bodies
that orbit other stars.  The asymmetry is the key discriminant here, and
is based on both the expected transit shapes
\citep{1999A&A...343..916L}, and those already discovered by
\citet{2018MNRAS.474.1453R}. Our method first searches for single transits,
and the asymmetry of these transits is then quantified. Most of the
artefacts that arise due to data anomalies and post-processing are
automatically rejected, but some manual inspection of a handful of light
curves is required to recover `true' events with confidence. For the
manual search, the false positive rate was fairly low (5/16 events were
classed as potential comet transits), but this success rate is very
likely to be lower if the search region was expanded.

Our search recovers the three deeper transits of KIC~3542116, and the
single transit of KIC~11084727, identified by
\citet{2018MNRAS.474.1453R}. However, without prior knowledge of these
events it is possible that one of these would have been missed because
our methods finds that the level of asymmetry is relatively low due to
the way we treat flagged data. We find two new events. KIC~3129239 shows
a reasonably symmetric transit, whose value of the asymmetry parameter
is relatively high, primarily because of a better fit near transit
center. The level of asymmetry is therefore not driven by slow egress,
and we disregard this event. KIC~8027456 shows an event very similar to
the previously known ones, albeit at a lower depth and somewhat longer
duration. We consider this event as a plausible cometary transit.

Considering the three stars that show plausible cometary transit events
in an H-R diagram suggests that they are younger than typical stars in
the \emph{Kepler} field, and have magnitudes and colours consistent with
stars approximately 100Myr old. This signature of youth is consistent
with our picture of how cometary source regions evolve; they deplete
over time so the probability of witnessing comet-related events towards
younger stars is presumably higher. Thus, their H-R diagram locations
give credence to the comet transit hypothesis for these events.

Given the detection of $\sim$10\% deep transit events towards
KIC~8462852, the detection of only a few more at 0.1\% levels suggests
that unless both our work and that of \citet{2018MNRAS.474.1453R} has
missed many events, searches for even shallower asymmetric events
\textbf{may} not yield significant numbers of new detections. While this
conclusion relies somewhat on the assumption that all of these events
share a similar origin, it seems that future searches are more likely to
be fruitful if they focus on the possibility of symmetric cometary
transits, and on searching among larger and younger samples of stars.

\section*{Acknowledgements}

We acknowledge valuable discussion with Tom Jacobs and Saul
Rappaport. GMK is supported by the Royal Society as a Royal Society
University Research Fellow. STH acknowledges support from an STFC
Consolidated Grant. This work used the \texttt{astropy}
\cite{2013A&A...558A..33A}, \texttt{matplotlib}
\cite{2007CSE.....9...90H}, \texttt{numpy} and \texttt{scipy}
\cite{2011arXiv1102.1523V} python modules. An online repository with
materials used in this work is available at
\href{https://github.com/drgmk/automated\_exocomet\_hunt}{https://github.com/drgmk/automated\_exocomet\_hunt}.













\bsp	
\label{lastpage}
\end{document}